\documentclass[conference]{IEEEtran}
\IEEEoverridecommandlockouts
\usepackage{cite}
\usepackage{amsmath,amssymb,amsfonts}
\usepackage{algorithmic}
\usepackage{graphicx}
\usepackage{textcomp}
\usepackage{amsmath}
\usepackage{booktabs} 
\usepackage{xcolor}
\usepackage{graphicx}
\usepackage{subcaption}
\usepackage{caption}
\usepackage{tikz}
\usepackage{soul, color, xcolor}
\usepackage{hyperref}
\hypersetup{
    colorlinks,
    citecolor=blue,
    filecolor=blue,
    linkcolor=blue,
    urlcolor=blue
}
\usepackage{amsmath, nccmath}
\usetikzlibrary{shapes,arrows,positioning}
\def\BibTeX{{\rm B\kern-.05em{\sc i\kern-.025em b}\kern-.08em
    T\kern-.1667em\lower.7ex\hbox{E}\kern-.125emX}}
    \usepackage[a4paper, top=1.9cm, bottom=2.54cm, left=1.3cm, right=1.3cm]{geometry}
\begin{document}

\title{A Novel Inverter Control Strategy with Power Decoupling for Microgrid Operations in Grid-Connected and Islanded Modes}
\author{\IEEEauthorblockN{\small{Yan Tong}{\textsuperscript{1}}}
\IEEEauthorblockA{\small\textit{1.Department of Electrical and Electronic} \\
\textit{\small{Engineering}}\\
\textit{\small{The Hong Kong Polytechnic University}}\\
\small{Hong Kong SAR} \\
\underline{\small{Stewart.tong@connect.polyu.hk}}}
\vspace{-1cm}
\and
\IEEEauthorblockN{\small{Qin Wang}{\textsuperscript{1}}}
\IEEEauthorblockA{\small\textit{1.Department of Electrical and Electronic} \\
\textit{\small{Engineering}}\\
\textit{\small{The Hong Kong Polytechnic University}}\\
\small{Hong Kong SAR} \\
\underline{\small{qin-ee.wang@polyu.edu.hk}}}
\vspace{-1cm}
\and
\IEEEauthorblockN{\small{Aihong Tang}{\textsuperscript{2}}}
\IEEEauthorblockA{\textit{\small{2.Department of Electrical Engineering}} \\
\textit{\small{School of Automation}}\\
\textit{\small{Wuhan University of Technology}}\\
\small{Wuhan, HuBei, China}\\
\underline{\small{tah@whut.edu.cn}}}
\vspace{-1cm}
}
\maketitle

\begin{abstract}

Grid-forming, particularly those utilizing droop control and virtual synchronous generators (VSG), can actively regulate the frequency and voltage of microgrid systems, exhibiting dynamic characteristics akin to those of synchronous generators. Although droop control and VSG control each have distinct benefits, neither can fully meet the diverse, dynamic needs of both grid-connected (GC) and islanded (IS) modes. Additionally, the coupling between active and reactive power can negatively impact microgrids' dynamic performance and stability. To solve these problems, this paper introduces a unified dynamic power coupling (UDC) model. This model's active power control loop can be tailored to meet diverse requirements. By implementing a well-designed control loop, the system can harness the advantages of both droop control and VSG control. In islanded mode, the proposed model can provide virtual inertia and damping properties, while in grid-connected mode, the inverter's active power output can follow the changed references without significant overshoot or oscillation. Furthermore, the model incorporates coupling compensation and virtual impedance based on a relative gain array in the frequency domain to facilitate quantitative analysis of power coupling characteristics. This paper outlines a distinct design process for the unified model. Finally, the proposed control method has been validated through simulation.
\end{abstract}


\section{Introduction}

\IEEEPARstart{T}{o} combat the growing effects of climate change and achieve significant reductions in carbon emissions, there is a rising global demand for renewable energy, particularly from clean sources like wind and solar power \cite{b1}. These renewable energy sources (RES) are valued for their sustainability, environmental benefits, diversity, and widespread availability. However, they are inherently variable and intermittent, as their availability depends heavily on weather and natural conditions \cite{b2}. Unlike traditional power generation methods, Renewable energy sources are integrated into the grid using power electronic inverters rather than synchronous generators. \cite{b3}. As the share of renewable energy in the grid increases, the system's inertia decreases, posing significant challenges for maintaining frequency regulation and stability. Renewable energy systems connect to the transmission network via a generation-side inverter, which optimizes generation efficiency, adjusts output voltage and current, and ensures compatibility with the grid's electrical characteristics \cite{b4}. The grid-side inverter further processes the energy output to align with the grid's frequency and voltage standards, facilitating smooth integration and enhancing the stability and reliability of the power system \cite{b5}.

To address these challenges, many studies focus on grid-side inverters, which can be controlled using two main strategies: Grid Following (GFL) and Grid Forming (GFM). Currently, most power electronic inverters in practical use adopt the GFL approach. This method employs a phase-locked loop (PLL) to continuously monitor the voltage at the point of common coupling (PCC), ensuring that the system aligns with the grid's voltage magnitude and phase \cite{b6}. In this setup, the GFL inverter operates similarly to a current source. However, to support a high penetration of GFL inverters, a significant number of Synchronous Generators (SGs) must be present in the grid. Without sufficient SGs, the reduced inertia makes it increasingly challenging to maintain voltage and frequency stability during disturbances. Researchers have proposed an advanced inverter control method known as GFM to improve grid stability and encourage renewable energy sources. GFM is widely applicable and can also be used in wireless charging\cite{b7}. and electric vehicle technologies\cite{b8}.. Inverters using the GFM approach exhibit characteristics similar to synchronous generators, allowing them to synchronize with the AC grid without relying on PLL \cite{b9}.

\begin{figure*}[t]
\centering
\includegraphics[width=6.6in]{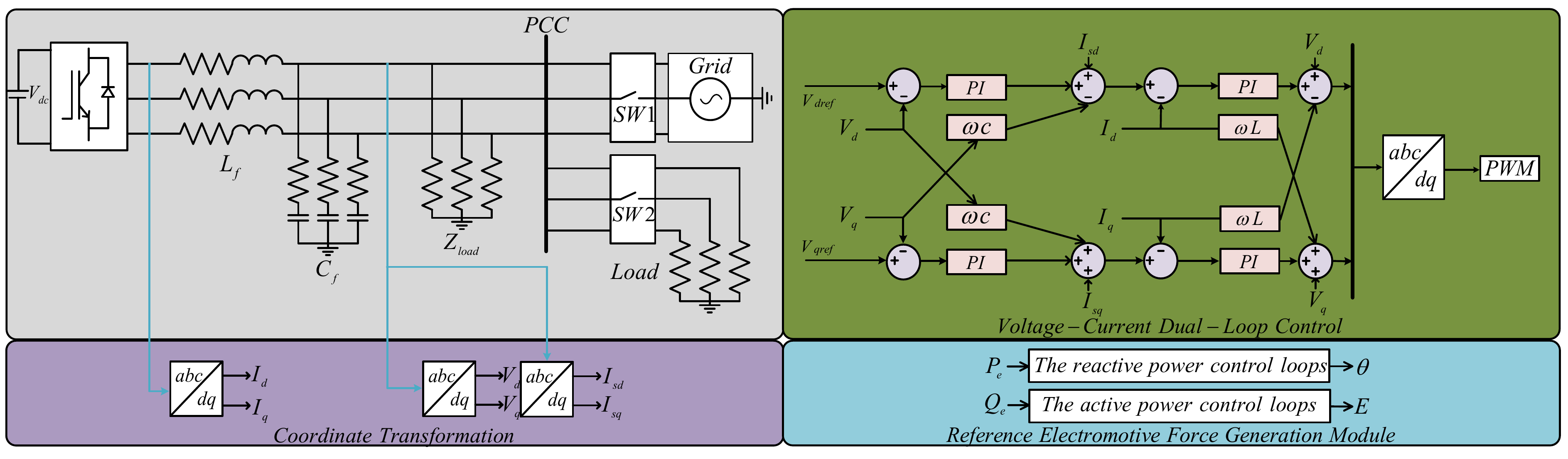} 
\caption{Main circuit structure of a grid-forming converter.}
\label{fig1}
\end{figure*}

The existing literature presents various advanced control strategies for managing phase angle, voltage magnitude, and power synchronization. These strategies include Droop Control (Voltage-Controlled) \cite{b10}, Virtual Synchronous Generator (VSG) Control \cite{b11}, Power Synchronization Control \cite{b12}, Synchronverter \cite{b13}, Matching  \cite{b14}, and Virtual Oscillator-based Control methods \cite{b15}. Through the above mentioned control methods, GFM inverters can actively regulate AC voltage and frequency and eliminate the instabilities associated with PLL. Their characteristics resemble those of a voltage source, providing greater precision and effectiveness compared to traditional GFL control. To date, research on GFM control has primarily focused on improving control strategies, conducting experimental simulations, and simplifying models\cite{b16}, while often overlooking practical engineering applications. In grid-connected (GC) mode, inverters utilizing VSG control usually exhibit overshoot and oscillations in output power. In islanded (IS) mode, the frequency variations of inverters with VSG control frequently do not comply with the Rate of Change of Frequency (ROCOF) requirements. Some studies propose generalized models but overlook the coupling of droop control \cite{b17}. In contrast, others \cite{b18} have proposed an Unfalsified Switching Adaptive Voltage Control that offers a novel approach to address this issue \cite{b19}.  Although the development of advanced control strategies effectively addresses various instabilities encountered during the integration of renewable energy sources into the grid, the diversity in inverter designs has led to challenges in standardization, increased costs, and limited applicability in real-world engineering contexts.

This paper makes three key contributions: 1) We propose a unified power model that can effectively operate in both droop and VSG control modes. It provides virtual inertia and damping in IS mode to meet ROCOF requirements while enabling fast and accurate active power output in GC mode without large overshoots or oscillations seen in traditional VSG control; 2) Building on the proposed unified power model, a unified dynamic power coupling (UDC) model is introduced, utilizing the relative gain array in the frequency domain. Compared to traditional droop control strategies, this model reduces differences in power regulation characteristics among Distributed Generation (DG) units by setting appropriate coupling compensation values. This approach minimizes phase angle differences between the power regulation curves of different DG units, significantly enhancing the system's flexibility and dynamic performance in a microgrid environment; 3) The impact of critical control parameters is analyzed, offering practical guidelines for designing the proposed model.

\section{Concept of dynamic  GFM control}
\subsection{Grid Integration Modelling}

When considering stability, traditional methods are insufficient. Fig.~\ref{fig1} illustrates the system's primary circuit, which includes coordinate transformation, reference electromotive force (EMF) generation, a virtual impedance loop, and dual closed-loop voltage and current control. This design focuses on modifying the reference EMF generation method, while other components remain essentially unchanged compared to traditional droop control, VSG control, and current improved methods. To develop a universal inverter control strategy applicable in both GC and IS modes, the following sections will introduce concepts in the order of traditional droop control, improved droop control, VSG control, and enhanced VSG control, leading to a new universal model. For analyzing inverter control in both GC and IS modes, a simulated grid module with switches and a simulated load module are connected at the PCC.

\subsection{Droop Control}

Droop control emulates the speed droop characteristics of synchronous generators by linking output power to system frequency. This allows automatic output adjustment to maintain stability in response to load changes without relying on the PLL. In droop control, output power is inversely proportional to frequency changes. For example, when the load increases and frequency decreases, the generator increases its output power. The control equation is as follows:
\begin{equation}
\omega  = {\omega _{ref}} - K_P^{droop}(P - {P_{ref}})
\label{q1}
\end{equation}
\begin{equation}
v = {v_{ref}} - K_d^{droop}(Q - {Q_{ref}})\label{q2}
\end{equation}
where \(\omega\) represents the measured angular frequency of the inverter's output voltage, \(v\) denotes the measured output voltage of the inverter, \(P\) is the active power output of the inverter, \(Q\) is the reactive power output, \(P_{ref}\) is the rated reference active power, \(Q_{ref}\) is the rated reference reactive power, \(\omega _{ref}\) is the rated reference angular frequency, \(v_{ref}\) is the reference voltage value,  \(K_P^{droop}\) is the droop coefficient for active power, and \(K_d^{droop}\) is the droop coefficient for reactive power.

\begin{figure}[htbp]
\centering
\includegraphics[width=0.45\textwidth]{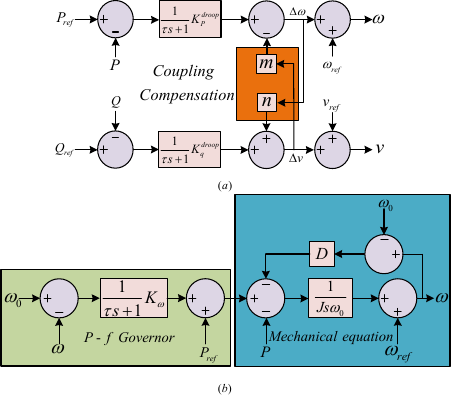}
\caption {Traditional Inverter Control Loop (a) Droop control,(b) Active power control loop of VSG.}
\label{2}
\end{figure}

When considering stability, it is inadequate to apply equations \eqref{q1} and \eqref{q2} exclusively to GFM power-voltage control. Establishing the relationship between frequency and power solely through the droop coefficient can often lead to the amplification of all oscillations in the measured power. To mitigate high-frequency harmonics in the measurements, a low-pass filter is typically combined with droop control. Additionally, In islanded microgrids, which are always in a dynamic state and experience power fluctuations, coupling compensation can be introduced to reduce the effects of uncontrollable power coupling. The concept is illustrated in Fig.~\ref{2}(a). The resulting transfer function is as follows:
\begin{equation}
\omega  = {\omega _{ref}} - \frac{1}{{\tau s + 1}}K_P^{droop}(P - {P_{ref}}) + m({v_{ref}} - v)\label{q3}
\end{equation}
\begin{equation}
v = {v_{ref}} - \frac{1}{{\tau s + 1}}K_d^{droop}(Q - {Q_{ref}}) - n({\omega _{ref}} - \omega )\label{q4}
\end{equation}
where \(\tau\) represents the time constant of the low-pass filter, which is typically defined as \(\tau  = 1/{\omega _c}\), where \(\omega _c\) is the cutoff angular frequency of the low-pass filter. Additionally, \(m\) and \(n\) are the voltage deviation compensation coefficient and the angular frequency deviation compensation coefficient, respectively.

\subsection{Virtual Synchronous Generator }
The concept of VSG is illustrated in Fig.~\ref{2}(b). Most VSG control strategies achieve this by utilizing short-term energy storage and appropriate control strategies to mimic the swing equations of synchronous generators. In these methods, the rotor inertia is typically the same, but the forms of damping characteristics differ. The \(P/f\) droop control regulator (5) and the mechanical equation of the synchronous machine (6) are expressed as follows:
\begin{equation}
{P_m} = {P_{ref}} + \frac{1}{{\tau s + 1}}{K_\omega }({\omega _{ref}} - \omega )\label{q5}
\end{equation}
\begin{equation}
\ddot \theta  = \frac{1}{J}({T_m} - {T_e} - {T_d})\label{q6}
\end{equation}
where \(T_m\) represents the mechanical torque, \(T_e\) denotes the electromagnetic torque, \(T_d\) is the damping torque, \(J\) refers to the moment of inertia, \(D\) is the damping coefficient, and \(P_m\) represents the mechanical power.
When the damping inertia is represented as \(D \cdot (\omega  - {\omega _0})\), \eqref{q6} is equivalent to
\begin{equation}
J\frac{{d\omega }}{{dt}} - J\frac{{d{\omega _0}}}{{dt}} = \frac{{{P_m} - P}}{\omega } - D(\omega  - {\omega _0})\label{q7}
\end{equation}
Combing \eqref{q5} and \eqref{q7}, we obtain
\begin{equation}
J\frac{{d\omega }}{{dt}} - J\frac{{d{\omega _0}}}{{dt}} = \frac{{{P_{ref}} + \frac{1}{{\tau s + 1}}{K_\omega }({\omega _{ref}} - \omega ) - P}}{\omega } - D(\omega  - {\omega _0})\label{q8}
\end{equation}
which is equivalent to
\begin{equation}
Js\omega  - Js{\omega _0} = \frac{{{P_{ref}} + \frac{1}{{\tau s + 1}}{K_\omega }({\omega _{ref}} - \omega ) - P}}{\omega } - D(\omega  - {\omega _0})\label{q9}
\end{equation}
The expression can be written in the form of a transfer function as follows:

\begin{equation}
\omega  = \frac{1}{J} \cdot \frac{1}{s} \cdot \frac{{{P_{ref}} + \frac{1}{{\tau s + 1}}{K_\omega }({\omega _{ref}} - \omega ) - P}}{\omega } - \frac{{D(\omega  - {\omega _0})}}{{Js}} + {\omega _0}\label{q10}
\end{equation}

{\begin{figure}[htbp]
\centering
\includegraphics[width=0.45\textwidth]{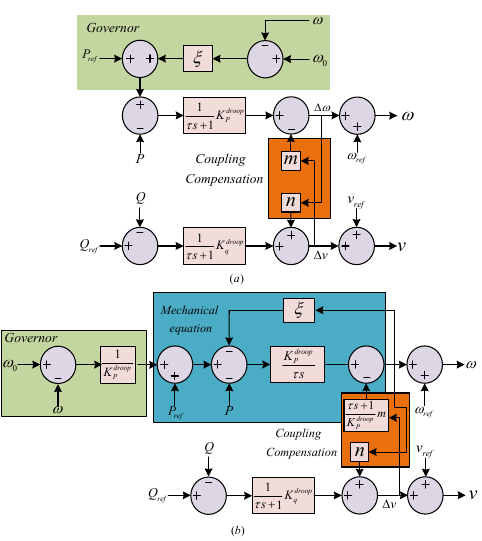}
\caption{Active power control loop of droop control with Governor}
\label{3}
\end{figure}}

\subsection{Unified Droop-VSG Control Strategy}
By expressing \(P_{ref}\) in terms of \(P_m\) and incorporating the \(P/f\) droop governor from the VSG control into the equation, the active reference power can be compensated for frequency variations, as illustrated in Fig.~\ref{3}(a). The corresponding control equations are then transformed as follows:
\begin{equation}
\omega  = {\omega _0} + \frac{1}{{\tau s + 1}}K_P^{droop}[{P_{ref}} + \xi ({\omega _{ref}} - \omega ) - P] + m({v_{ref}} - v)\label{q11}
\end{equation}

\noindent where \(\xi\) represents the adjustment coefficient in the governor control and \(\omega _{ref}\) denotes the reference frequency for the governor control.
The equation \eqref{q11}) can be rewritten as follows:

\begin{equation}
\begin{aligned}
{P_{ref}} + \xi &({\omega _{ref}} - \omega ) - P - \frac{1}{{K_P^{droop}}}(\omega  - {\omega _0})\\
&{\rm{ + }}\frac{{\tau s + 1}}{{K_P^{droop}}} \cdot m \cdot ({v_{ref}} - v){\rm{ = }}\frac{{\tau s}}{{K_P^{droop}}}(\omega  - {\omega _0})\label{q12}
\end{aligned}
\end{equation}

By comparing equations \eqref{q9} and \eqref{q12}, it can be observed that the introduction of the \(P/f\) droop governor results in both droop control and VSG control sharing a common active power control loop under the condition when

\begin{equation}
\begin{aligned}
D & = \xi \quad & J \omega_0 & = \frac{\tau s}{K_P^{droop}} \quad & \frac{1}{\tau s + 1} K_\omega & = \frac{1}{K_P^{droop}} \\
\end{aligned}
\label{q13}
\end{equation}

The droop active power control in equation \eqref{q12} is a special case in equation \eqref{q9}, with further details substantiated in \cite{b1}. In the steady state of the system, the active and reactive power of the grid can be expressed as:
\begin{equation}
{P_r} = \frac{{EV\sin \theta }}{X}\label{q14}
\end{equation}
\begin{equation}
{Q_r} = \frac{{{E^2} - EV\cos \theta }}{X}\label{q15}
\end{equation}
where \(P_r\) and \(Q_r\) are active and reactive power at the grid side, \(E\) and \(V\) represent the voltage values of the grid and the inverter, \(\theta\) is the phase angle difference between the inverter input voltage and the grid-side voltage, and \(X\) denotes the line impedance. 

For active power control design, \(E\) and \(V\) are assumed constants. \eqref{q14} is linearized as:
\begin{equation}
{P_r} \approx  \frac{{{E_0} \cdot {V_0} \cdot \theta }}{X} \label{q16}
\end{equation}

\begin{figure}[htbp]
\centering
\includegraphics[width=0.45\textwidth]{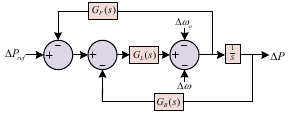}
\caption{Simplified small-signal model of UDC}
\label{4}
\end{figure}

The small-signal representation of the active control loop in the UDC model, based on a comparison of Droop control and VSG control, can be simplified as shown in Fig.~\ref{4}. In this model, the loop dynamics are controlled using a feedforward function, a feedback function, and a filtering function.

\section{Experimental Simulation and Parameter Design}

A simulation model was developed in Simulink to validate the proposed UDC model, as illustrated in Fig.~\ref{fig1} and the main circuit parameters are shown in Table \ref{tab1}. An ideal DC voltage source is connected to the input of the GFM inverter to simulate a stable input voltage provided by the generation side. This is connected to the grid connection point via a transmission line equipped with an LC filter. On the grid side, a three-phase ideal voltage source simulates the grid, while a load with switches simulates load variations. When Switch 1 is closed and Switch 2 is open, the system reference power is adjusted to simulate changes in reference power under GC mode. Conversely, when Switch 1 is open, the state of Switch 2 is toggled to simulate load power variations in IS mode.

\begin{table}[htbp]
\captionsetup{justification=centering, singlelinecheck=false}
\caption{Characteristic parameters of the system}
\centering
\renewcommand\arraystretch{1.4}
\setlength{\tabcolsep}{1.8mm}{
\begin{tabular}{cc|cc}
\hline
\hline
Parameter & Value & Parameter & Value \\
\hline
DC voltage $V_{DC} $ & 1500 V & Rated Voltage $U_{n} $ & 380 V \\
Filter inductor $L$ & 3 mH & Nominal Voltage $U_{b} $ & 6000 V \\
Filter capacitor $C_{f} $ & 35 uF & Rated Capacity $S_{b} $ & 800 kVA \\
Passive damping  $R$ & 0.5 $\Omega $ & Reference Power $P_{f}$ & 12 KW \\
Switching frequency $f$ & 20 kHz & Load Power $P_{Load} $ & 1 KW \\
Virtual inertia $J$ & 3.36 kg·m² & Damping coefficient $D$ & 100  \\
\hline
\hline
\end{tabular}}
\label{tab1}
\end{table}

In GC mode, the closed-loop transfer function should be expressed as:
\begin{equation}
\cfrac{\triangle P}{\Delta P_{r} } = \frac{V_{0}V_{g}G_{L}(s)   }{sX+sG_{F}(s)G_{L}(s)+V_{0}V_{g}G_{L}(s)G_{B}(s)  }  \label{q17}
\end{equation}
In IS mode, using $\Delta \omega $ and $\Delta P_{Load} $ to represent the variations between frequency and load, the closed-loop transfer function at this time can be expressed as:
\begin{equation}
\cfrac{\triangle \omega }{\Delta P_{load} } = \frac{-G_{B}(s)G_{L}(s)   }{1+G_{F}(s)G_{L}(s)}  \label{q18}
\end{equation}
It was noted that the $G_{F} (s)$ employs a proportional controller to ensure the stability of the system frequency. This can be combined with the deviation controller representing the filter:
\begin{equation}
G(s) = \cfrac{G_{F} (s)}{1+G_{L} (s)G_{F} (s)} \label{q19}
\end{equation}

To ensure that the inverter possesses power-sharing capability in IS mode and meets the ROCOF requirement when \(P_L\) changes, thereby enabling the inverter's angular frequency to adapt accordingly, the system's transfer function should incorporate at least three negative real poles. These poles serve to provide virtual inertia, guaranteeing that the transfer function constitutes a super-damped second-order system, which in turn ensures that the angular frequency remains free from overshoot. Similarly, to enable more accurate tracking of power changes when \(P_r\) varies in GC mode, the transfer function should include a negative real zero. The transfer function is as follows:
\begin{equation}
G(s) =\cfrac{1 + sT_{z1} }{(1 + sT_{p1})(1 + sT_{p2})(1 + sT_{p3})} \label{q20}
\end{equation}

This transfer function structure ensures the desired characteristics for both the IS and GC modes of operation, providing virtual inertia and accurate power tracking capabilities to the inverter system. The corresponding small-signal transfer function models for GC mode and IS mode are as follows:

\begin{equation}
\begin{medsize}
\begin{aligned}
&\cfrac{\triangle P}{\Delta P_{r} } = \cfrac{V_{0}V_{g}(1 + sT_{z1})}{sX(1 + sT_{p1})(1 + sT_{p2})(1 + sT_{p3})+ V_{0}V_{g}(1 + sT_{z1})(\beta s+1)}\label{q21}
\end{aligned}
\end{medsize}
\end{equation}

\begin{figure}[htbp]
    \centering
    \begin{minipage}{0.45\textwidth}
        \centering
        \includegraphics[width=\linewidth]{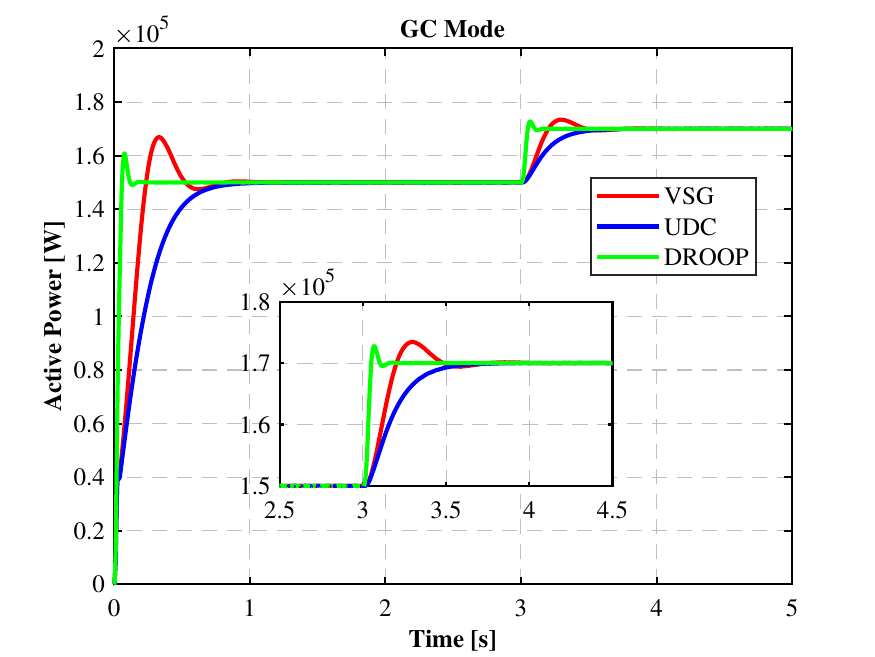}
        \caption{Simulation Results of Step Responses for VSG, UDC, and Droop Control under Step Change in Reference Power}
        \label{5}
    \end{minipage}
    \hfill
    \begin{minipage}{0.45\textwidth}
        \centering
        \includegraphics[width=\linewidth]{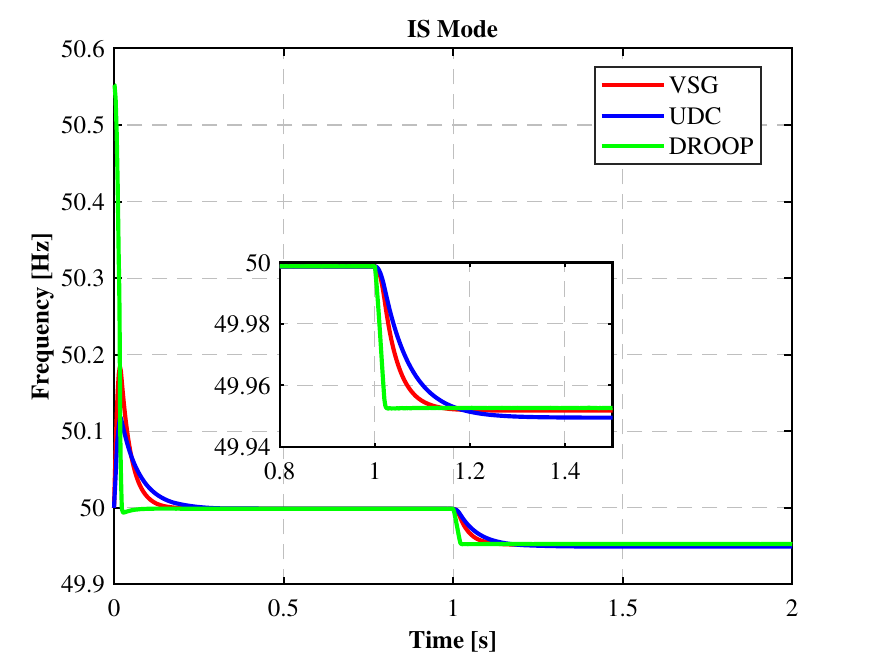}
        \caption{Simulation Results of Step Responses for VSG, UDC, and Droop Control under Step Change in Load Power}
        \label{6}
    \end{minipage}
    \label{fig:comparison}
\end{figure}

The three control strategies were individually implemented in the primary system and simulated using Simulink software. In the GC microgrid, when the active reference power $P_{ref} $ undergoes a step change, the output power of the AC grid responds accordingly, as illustrated in Fig.~\ref{5}. Compared to VSG control, the proposed inverter reduces the active power overshoot to 5\%, and the settling time is shortened from 3.8 seconds to 3.5 seconds.Although this method reduces the rise time, it significantly improves the stability of the system.

\begin{table}[htbp]
\caption{Comparison for Different Control Methods in GC mode}
\centering
\renewcommand{\arraystretch}{1.2} 
\setlength{\tabcolsep}{6pt} 
\begin{tabular}{@{}lccc@{}}
\toprule
\textbf{Control Method} & \textbf{Rise Time (s)} & \textbf{Settling Time (s)} & \textbf{Overshoot (\%)} \\ \midrule
Droop Control           & 3.0                   & 3.1                        & 10                     \\
VSG Control             & 3.0                   & 3.8                        & 13                     \\
UDC Control           & 3.0                   & 3.5                        & \textbf{5}             \\ \bottomrule
\end{tabular}
\label{table:rise_settling_time}
\end{table}

Similarly, in IS mode, a step change in load power was achieved by closing the switch at 1 second. The simulation results are shown in Fig.~\ref{6}. Compared to traditional VSG and droop control, the proposed inverter demonstrates superior frequency tracking performance, ensuring that the system provides frequency support during load transients and meets the ROCOF requirements.

In an IS microgrid, the system is constantly in a dynamic state, experiencing power fluctuations. Compared to traditional droop control strategies, the improved droop control strategy can significantly reduce the differences in power regulation characteristics among various units by setting appropriate coupling compensation values. This approach greatly minimizes the phase angle differences in the power regulation curves of the units, thereby enhancing the degrees of freedom in droop control and resulting in significantly better dynamic performance.

\section{Conclusion}

As the share of RES in power systems continues to increase, GFM control can flexibly adjust the power-frequency characteristic curve according to actual requirements, actively managing the system's frequency and voltage. Currently, most industrial applications still utilize GFL control. However, the varying demands for GFM control across different scenarios lead to significant costs.

In this study, we have examined the limitations of traditional VSG and droop control strategies. By analyzing their small-signal transfer functions and considering dynamic power fluctuations in microgrids, we develop a generalized model with coupling compensation values. This model effectively integrates VSG and droop control characteristics, which can enhance the microgrid's dynamic performance across diverse scenarios. The findings provide valuable insights for industrial applications, especially in localized microgrids prone to grid disconnection due to extreme conditions. Under fault conditions such as voltage sag or three-phase short circuit faults, the proposed UDC control method still retains the virtual inertia and damping characteristics of the VSG method, providing a certain level of resilience to faults. Theoretically, it can effectively mitigate transient instability. Due to the decentralized nature of GFM, it can utilize signals such as bus frequency and voltage amplitude for distributed control. In future research, we will explore the scalability of this model in large-scale power grids and investigate whether its implementation has any impact on grid performance. Additionally, we will incorporate experiments to estimate processing time and verify the feasibility of the proposed approach.

\end{document}